\documentclass[twocolumn, showpacs, nofootinbib, aps, prd]{revtex4}

\usepackage{graphicx}
\usepackage{dcolumn}
\usepackage{bm}
\usepackage{hyperref}
\usepackage{color}

\begin{document}

\title{Finding a way to determine the pion distribution amplitude from the experimental data}
\author{Tao Huang$^1$}
\email{huangtao@ihep.ac.cn}
\author{Xing-Gang Wu$^{2}$}
\email{wuxg@cqu.edu.cn}
\author{Tao Zhong$^{1}$}
\email{zhongtao@ihep.ac.cn}

\address{$^1$ Institute of High Energy Physics and Theoretical Physics Center for Science Facilities, Chinese Academy of Sciences, Beijing 100049, P.R. China \\
$^{2}$ Department of Physics, Chongqing University, Chongqing 401331, P.R. China}

\date{\today}

\begin{abstract}

It is believed that one can extract more accurate information of the pion distribution amplitude from the pion-photon transition form factor (TFF) due to the single pion in this process. However the BABAR and Belle data of the pion-photon TFF have a big difference for $Q^2\in [15,40]$~GeV$^2$, and at present, the pion DA can not be definitely determined from the pion-photon TFF. it is crucial to find the right pion DA behavior and to determine which data is more reliable. In this letter, we perform a combined analysis of the most existing data of the processes involving pion by using a general model for the pion wavefunction/DA. Such a DA model can mimic all the existed pion DA behaviors, whose parameters can be fixed by the constraints from the processes $\pi^0\to\gamma\gamma$, $\pi\to\mu\nu$, $B\to\pi l \nu$, and etc. Especially, we examine the $B \rightarrow \pi$ transition form factors that provides another constraint to the parameter $B$ in our DA model, which results in $B \in[0.00,0.29]$. This inversely shows that the predicted curve for the pion-photon TFF is between the BABAR and Belle data in the region $Q^2\in$ $[15,40]$~GeV$^2$. It will be tested by coming more accurate data at large $Q^2$ region, and the definite behavior of pion DA can be concluded finally. \\

\end{abstract}


\maketitle

The recent data by the Belle Collaboration~\cite{belle} is dramatically different from those reported by the BABAR Collaboration~\cite{babar}, i.e. instead of a pronounced growth of the TFF at high $Q^2$ region, observed by BABAR, the Belle data are compatible with the well-known asymptotic prediction $Q^2 F_{\pi \gamma} (Q^2)|_{Q^2\to\infty} \to \sqrt{2}f_\pi$, where the pion decay constant $f_{\pi}=130.41 \pm 0.03\pm0.20$ MeV~\cite{fpi}. One thinks that it would be a big challenge to QCD if the the BABAR data are confirmed. One should ask a question what is the realistic light-cone wave function of the pion. Theoretically the pion-photon TFF depends heavily on the pion DA and sensitive to it in the higher energy region. Such large discrepancy between the Belle and the BABAR data at the high $Q^2$ region, shows that the pion-photon TFF can't discriminate the different pion DA models. On other hand, All the processes involving pion can provide more strong constraints on the pion DA for determining it. It would be helpful to have a general pion DA model that can mimic all the DA behaviors suggested in the literature.

Recently, we have raised a pion wave function (WF)~\cite{spatialWF1,spatialWF2} following the idea of Refs.\cite{spin1,spin2,spin3}, in which the BHL prescription~\cite{bhl} and the Melosh rotation~\cite{melosh} are applied. More explicitly, the full form of the pion WF is written as
\begin{equation}\label{wave}
\Psi(x,{\bf k}_{\perp})=\sum_{\lambda_{1}\lambda_{2}} \chi^{\lambda_{1} \lambda_{2}}(x,{\bf k}_{\perp}) \Psi^{R}(x,{\bf k}_{\perp}) ,
\end{equation}
where $\chi^{\lambda_{1} \lambda_{2}}(x,{\bf k}_{\perp})$ stands for the spin-space WF and the spatial WF
\begin{equation}\label{wave1}
\Psi^{R}(x,{\bf k}_{\perp})=A\varphi_\pi(x)\exp\left[-\frac{{\bf k}_{\perp}^2 +m_q^2}
{8{\beta}^2x(1-x)}\right].
\end{equation}
The pion DA can be derived by integrating out the transverse momentum dependence of the WF, and we obtain
\begin{widetext}
\begin{eqnarray}
\phi_\pi(x,\mu_0^2) = \frac{\sqrt{3}A m_q \beta}{2\pi^{3/2}f_\pi} \sqrt{x(1-x)} \varphi_\pi(x) \times \left( \mathrm{Erf} \left[\sqrt{\frac{m_q^2+\mu_0^2}{8\beta^2 x(1-x)}}\right]- \mathrm{Erf } \left[ \sqrt{\frac{m_q^2}{8\beta^2 x(1-x)}}\right] \right), \label{phimodel}
\end{eqnarray}
\end{widetext}
where $\mathrm{Erf}(x)=\frac{2}{\sqrt{\pi}} \int_0^x e^{-t^2}dt$, $\mu_0$ stands for the factorization scale. The pion DA at any scale $\mu^2$ can be derived from the initial DA $\phi_\pi(x,\mu_0^2)$ through QCD evolution~\cite{lb}. The function $\varphi_\pi(x)$ dominates the longitudinal distribution broadness of the WF, which can be expanded in the Gegenbauer polynomials, and by keeping its first two terms, we have $\varphi_\pi(x) =\left[1+B\times C^{3/2}_2(2x-1)\right]$.

In the above model, there are four parameters $A$, $\beta$, $m_q$ and $B$. All the processes involving pion can provide strong constraints on the pion DA. Then these parameters can be fixed by a combination of various constraints,
\begin{itemize}
\item The process $\pi \rightarrow \mu\nu$ provides the WF normalization condition~\cite{bhl},
    \begin{equation}
      \int^1_0 dx \int \frac{d^{2}{\bf k}_{\perp}}{16\pi^3}\Psi(x,{\bf k}_{\perp}) ={f_{\pi}}/{2\sqrt{6}} .
    \end{equation}
\item The process $\pi^0\rightarrow\gamma\gamma$ provides a sum rule for the WF~\cite{bhl},
    \begin{equation}
      \int^1_0 dx \Psi(x,{\bf k}_{\perp}=0)={\sqrt{6}}/{f_{\pi}} .
    \end{equation}
\item The requirement of a reasonable value for the light valence quark mass $m_q$ provides another constraint, whose value should be around $0.30$ GeV.
\item The parameter $B$ can be regarded as an effective parameter that absorbs the contributions from all higher Gegenbauer polynomials, which dominantly determines the broadness of the WF. Its value can be fixed by high energy processes involving pion.
\end{itemize}

\begin{table}[htb]
\caption{Typical pion WF parameters under the condition of $m_q=0.30$ GeV, and its corresponding Gegenbauer moments.}
\begin{center}
\begin{tabular}{|c||c|c||c|c|c|}
\hline ~~~$B$~~~ & ~$A ({\rm GeV}^{-1})$~& ~$\beta ({\rm GeV})$~ & ~$a_2(\mu_0^2)$~ & ~$a_4(\mu_0^2)$~  \\
\hline
~$0.00$~ & ~$25.01$~& ~$0.586$~ & ~$0.027$~ & ~$-0.027$~  \\
\hline
~$0.30$~ & ~$20.22$~& ~$0.668$~ & ~$0.362$~ & ~$-0.018$~  \\
\hline
~$0.60$~ & ~$16.59$~& ~$0.745$~ & ~$0.679$~ & ~$0.020$~   \\
\hline
\end{tabular}
\label{tab1}
\end{center}
\end{table}

It is obvious that one can determine the right pion DA behavior by comparing the theoretical predictions derived by using different pion DA with the experimental data. The pion DA as shown by Eq.(\ref{phimodel}) can be conveniently adopted for such purpose. We compare this model with other existed models in the literature and find it can mimic all the existed pion DA behaviors, e.g. as shown in Table \ref{tab1}, keeping $m_q=0.30$ GeV and varying the parameter $B$ within the region of $[0.00,0.60]$, the pion DA (\ref{phimodel}) shall vary from asymptotic-like to CZ-like. A detailed discussion on this point can be found in Refs.\cite{spatialWF1,spatialWF2}. Then, to determine the right DA behavior is equivalent to find a proper $B$ that can well explain the data.

The process $\gamma\gamma^*\to \pi^0$ provides the simplest example of the perturbative application to exclusive processes since this process relates two photons with a single pion. It is usually regarded as a suitable platform to determine the pion DA. However, the large discrepancy between the Belle and the BABAR data at the high $Q^2$ region doesn't help to discriminate the different pion DA models or to fix the parameter B. In fact, by varying $B$ within the region of $[0.00,0.60]$, it has been shown that the estimated form factors at large $Q^2$ region are quite different and they are in between the BABAR and Belle data for various value of $B$~\cite{spatialWF1,spatialWF2,spatialWF3}.

In the literature, the $B \to \pi$ transition form factor $f^{B\to\pi}_{+}$ is usually adopted to extract the CKM matrix element $|V_{ub}|$. A consistent analysis of $f^{B\to\pi}_{+}$ within the whole physical region has been in Ref.\cite{huangwu2004}. In the framework of QCD light-cone sum rules, the expression for $f^{B\to\pi}_{+}$ heavy depends on the correlation function: different choice of the currents in the correlation function shall result in different expressions, in which, the pionic different twist structures provide different contributions. More explicitly, as a direct estimation up to twist-4 level, the $f^{B\to\pi}_{+}$ contains the full pion twist-2, 3 and 4 contributions by taking the conventional correlator~\cite{BPIT234_95,BPIT234_98,BPIT234_01,BPIT234_05,BPIT234_08}. On the other hand, by taking the correlator with a chiral current~\cite{BPIT2_01,BPIT2_03,BPIT2_08,BPIT2_09,BPIT2_12}, the most uncertain twist-3 contributions can be eliminated; while by taking the correlator suggested by Refs.\cite{BPIT3_04,BPIT3_11}, only the twist-3 contributions are kept up to twist-4 level.

In Refs.\cite{BPIT2_09,BPIT2_12}, it has been pointed out that the pionic leading twist contribution is dominant, which provides a greater than $95\%$ contribution to the form factors. This provides us a good change to study the properties of the leading twist pion DA, that is, by using the experimental values for $|V_{ub}|$ and $f^{B\to\pi}_{+}(0)|V_{ub}|$, one can expect to obtain a strong constraint for the parameter $B$.

On the experimental side, from the measured shapes of the form factor for $B \to \pi l \tilde{\nu}$, the CKM matrix element $|V_{ub}|$ multiplied by $f^{B \to \pi}_+(0)$ is~\cite{FVUB}:
\begin{eqnarray}
f^{B \to \pi}_+(0) |V_{ub}| = (9.4 \pm 0.3 \pm 0.3) \times 10^{-4} . \label{fvub}
\end{eqnarray}
The CKM matrix element $|V_{ub}|$ is $(3.23\pm0.31) \times 10^{-3}$, which is derived from a simultaneous fit to the experimental partial rates and lattice points on the exclusive process $B \to \pi l \tilde{\nu}$ versus $q^2$~\cite{lattice}. As a combination, we obtain an moderate experimental limit for $f^{B \to \pi}_+(0)$, i.e.
\begin{equation}
f^{B \to \pi}_+(0)=0.291^{+0.010}_{-0.009}.
\end{equation}
It is noted that this upper limit $f^{B \to \pi}_+(0)= 0.301$ agrees with the theoretical limit derived in most of the literature, either under the pQCD approach (cf. $f^{B \to \pi}_+(0)=0.265\pm0.032$~\cite{huangwu2004}) or the QCD light-cone sum rules approach (cf. $f^{B \to \pi}_+(0)=0.258\pm0.031$~\cite{BPIT234_05}) or the extrapolated lattice QCD approach (cf. $f^{B \to \pi}_+(0)=0.27 \pm0.11$~\cite{latticebpi}). While the lowest limit $0.282$ for $f^{B \to \pi}_+(0)$ is somewhat too strict, however a better determination on this value is less important here, since it only determines the lower limit for the parameter $B$, which already reaches its lowest value $B_{\rm min}=0.00$ for the present choice.

\begin{figure}[t]
\centering
\includegraphics[width=0.45\textwidth]{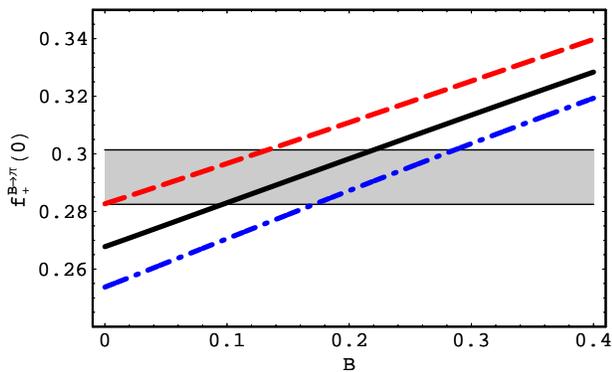}
\caption{The value of $f^{B\to\pi}_+(0)$ versus the parameter $B$. The solid, the dashed and the dash-dot lines stand for central, the upper and the lower values for $f^{B\to\pi}_+(0)$ that are derived from the QCD light-cone sum rules~\cite{BPIT2_12}. The shaded band represents the value derived from a combination of the data on $f^{B\to\pi}_+(0)|V_{ub}|$ and $|V_{ub}|$~\cite{FVUB,lattice}. } \label{FBPI}
\end{figure}

On the theoretical side, we adopt the light-cone sum rules derived in Ref.\cite{BPIT2_12} for our analysis. During the calculation, we take the present DA model (\ref{phimodel}) to do the calculation; while the other input parameters are chosen to be the same as those of Ref.\cite{BPIT2_12}. We present the value of $f^{B\to\pi}_+(0)$ versus the parameter $B$ in Fig.(\ref{FBPI}), where the theoretical errors are the squared average of those dominant errors caused by taking the $\overline{\rm MS}$ $b$-quark mass $\bar{m}_b(\bar{m}_b)=4.164\pm0.025$ GeV~\cite{mb}, the $B$ meson decay constant up to next-to-leading order $f_B=214^{+7}_{-5}$~\cite{BPIT234_08} and the effective threshold $s^B_0=(34\pm0.5)\; {\rm GeV}^2$. We obtain the range of the parameter $B$:
\begin{eqnarray}
B = [0.00, 0.29] .  \label{bvub}
\end{eqnarray}

\begin{figure}[ht]
\centering
\includegraphics[width=0.4\textwidth]{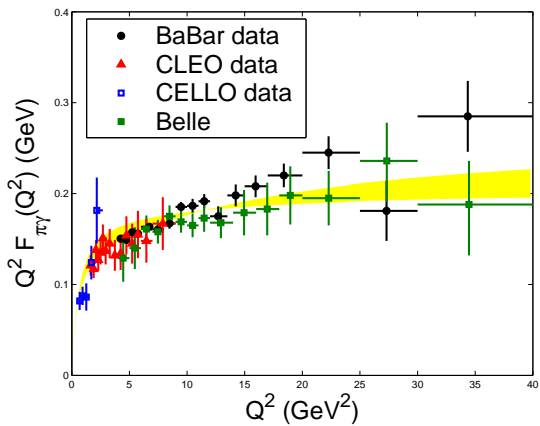}
\caption{$Q^2 F_{\pi\gamma}(Q^2)$ by taking $m_q=0.30$ GeV and $B\in[0.00,0.29]$, where as a comparison, the Belle~\cite{belle}, BABAR~\cite{babar}, CELLO~\cite{CELLO} and CLEO~\cite{CLEO} data are presented. } \label{figpionff}
\end{figure}

As a summary, we have performed a combined analysis of the most existing data of the processes involving pion by using a flexible model for the pion DA. Such a DA model can mimic all the existed pion DA behaviors, whose parameters can be fixed by the constraints from the processes $\pi^0\to\gamma\gamma$, $\pi\to\mu\bar{\nu}$, $B\to\pi l\bar{\nu}$, and etc. Especially, the experimental data on the $B \to \pi$ transition form factor provides useful information for determining the pion DA.

Using the experiments on the values of $f^{B\to\pi}_+$ at $q^2=0 \;GeV^2$, we show that the reasonable range of the effective parameter $B$ of pion DA is $[0.00,0.29]$. It is noted that such region is consistent with Brodsky and Teramond's holographic DA model whose second Gegenbauler moment is about $0.145$~\cite{holo}. Therefore all of parameters in our DA model are fixed to a certain confidence level. A better understanding of other pionic processes, such as $D\to\pi l\bar{\nu}$, shall provide further constraints on the wave function parameters.

The pion-photon TFF can be calculated exactly by using the pion DA light-cone wave function (\ref{phimodel}). In the small $Q^2$ region, $Q^2\lesssim 15~GeV^2$, the pion-photon TFF for $B\sim [0.00,0.29]$ can explain the Belle~\cite{belle}, BABAR~\cite{babar}, CELLO~\cite{CELLO} and CLEO~\cite{CLEO} experimental data. While for the large $Q^2$ region, as can be seen from Fig.(\ref{figpionff}), the pion-photon TFF is sensitive to the parameter $B$. The result with $B\in[0.00,0.29]$ is consistent with previous observation that a moderate pseudo-scalar ($\pi$, $\eta$ or $\eta'$) DA $B\sim 0.1-0.3$ that is close to asymptotic-like can roughly explain the TFFs $Q^2 F_{\pi\gamma}$, $Q^2 F_{\eta\gamma}$ and $Q^2 F_{\eta'\gamma}$ data simultaneously~\cite{spatialWF2,etagamma,stan}. It has shown from Fig.(\ref{figpionff}) that the curve with the parameter $B=0.00$ is more close to the Belle data and the curve with the parameter $B=0.29$ is between the BABAR and Belle experimental data. Thus, it will be tested by coming more accurate data at large $Q^2$ region, and the definite behavior of pion DA can be concluded finally.

{\bf Acknowledgments}: The authors would like to thank Z.H. Li and N. Zhu for helpful discussions. This work was supported in part by Natural Science Foundation of China under Grant No.10975144 and No.11075225, and by the Program for New Century Excellent Talents in University under Grant NO.NCET-10-0882.

\end{document}